\RequirePackage{etoolbox}
\csdef{input@path}{%
 {sty/}
 {img/}
}%
\csgdef{bibdir}{bib/}

\documentclass[ba]{imsart}
\pubyear{0000}
\volume{00}
\issue{0}
\doi{0000}
\firstpage{1}
\lastpage{1}

\usepackage{amsthm}
\usepackage{amsmath}
\usepackage{natbib}
\usepackage{graphicx}
\usepackage{amssymb,pdflscape,subcaption} 

\startlocaldefs
\newcommand{\bz}{\boldsymbol{Z}}
\newcommand{\bx}{\boldsymbol{X}}
\newcommand{\bu}{\boldsymbol{u}}
\newcommand{\logit}{\mbox{logit}}

\newcommand{\btau}{\boldsymbol\tau}
\newcommand{\br}{\boldsymbol{r}}
\newcommand{\bs}{\boldsymbol{s}}
\newcommand{\bbeta}{\boldsymbol\beta}

\newcommand{\bmu}{\boldsymbol\mu}

\newcommand{\cx}{{\cal X}}
\newcommand{\cz}{{\cal Z}}
\newcommand{\Prob}{\mathbb{P}}

\newcommand{\notimplies}{%
  \mathrel{{\ooalign{\hidewidth$\not\phantom{=}$\hidewidth\cr$\implies$}}}}
\endlocaldefs

\begin{document}


\begin{frontmatter}
\title{Latent Space Approaches to Community Detection in Dynamic Networks}

\runtitle{Latent Space Approaches to Community Detection in Dynamic Networks}

\begin{aug}
\author{\fnms{Daniel K.} \snm{Sewell}\thanksref{addr1}\ead[label=e1]{daniel-sewell@uiowa.edu}}
\and
\author{\fnms{Yuguo} \snm{Chen}\thanksref{addr2}\ead[label=e2]{yuguo@illinois.edu}}

\runauthor{Sewell and Chen}

\address[addr1]{Department of Biostatistics, University of Iowa, Iowa City, IA 52242, \printead{e1}}
\address[addr2]{Department of Statistics, University of Illinois at Urbana-Champaign, Champaign, IL 61820, \printead{e2}}

\end{aug}

\begin{abstract}
Embedding dyadic data into a latent space has long been a popular approach to modeling networks of all kinds.  While clustering has been done using this approach for static networks, this paper gives two methods of community detection within dynamic network data, building upon the distance and projection models previously proposed in the literature.  Our proposed approaches capture the time-varying aspect of the data, can model directed or undirected edges, inherently incorporate transitivity and account for each actor's individual propensity to form edges.  We provide Bayesian estimation algorithms, and apply these methods to a ranked dynamic friendship network and world export/import data.
\end{abstract}


\end{frontmatter}


\section{Introduction}
Researchers are often interested in detecting communities within dyadic data.  These dyadic data are represented as networks with a certain number of actors which can form amongst themselves relationships/connections called edges.  Some examples of such data include social networks, collaboration networks, biological networks, food-webs, power grids, linguistic networks.  These dyadic data can have directed or undirected edges, have zero-one or weighted edges and can come in the form of static or dynamic (time-varying) networks.  Clustering these data into communities can lead to better understanding of the organization of the objects in the network, and, for dynamic networks, how this organization evolves over time.

\cite{xing2010state} developed a dynamic mixed membership stochastic blockmodel.  This work builds on the stochastic blockmodel \citep{holland1983stochastic}, further developed into the mixed membership blockmodel \citep{airoldi2005latent}.  In the work of \cite{xing2010state}, each actor has an individual membership probability (time-varying) vector and, based on this probability vector, can choose certain roles with which to interact with other actors.  A different approach to be taken in this paper begins with the work of \cite{hoff2002latent} where the actors are embedded either within a latent Euclidean space, referred to as the distance model, or within a hypersphere, referred to as the projection model.  \cite{handcock2007model} used their distance model and performed community detection on the latent actor positions.  Further, the distance model of \cite{hoff2002latent} was extended by \cite{sewell2014latent}
and \cite{durante2014nonparametric} to include dynamic network data, and \cite{sewell2014analysis,sewell2014weighted} extended their dynamic model to allow for various types of weighted edges.  

Applying a latent space model has distinct advantages over other approaches, such as blockmodeling.  Using a latent space approach allows the user to capture local and global structures.  The output yields meaningful visualization of the data, providing rich qualitative information.  Transitivity and reciprocity, two important features of many networks, is inherently incorporated in the model.  In our proposed methodology, the variation in individual edge propensities, often described by their degree distributions, is accounted for.  Finally, homophily can be easily incorporated into the model just as in latent space approaches for static networks.  That is, exogenous actor attributes can be incorporated into the linear modeling; these covariates may also be added by extending the hierarchical model to predict cluster assignments \citep[see][]{gormley2010mixture}.

This work provides advances beyond the existing literature on latent space network models by constructing mechanisms to perform community detection on dynamic network data and providing Bayesian estimation methods.  Specifically, the primary goals of our proposed methodology are to determine what communities exist in the network, which actors belong to these communities and how these actors change communities over time.  The proposed methodology accomplishes these clustering goals while maintaining a very flexible framework that can handle directed or undirected dyads and virtually any type of weighted edges, e.g., ranked dynamic network data.  Information is borrowed across time to obtain more accurate clustering estimates.  In addition, we present clustering models based on the two common geometries used in the latent space literature, Euclidean spaces and hyperspheres.  To the authors' knowledge there is no existing latent space methodology that achieves these community detection goals for dynamic networks with either geometry, and no such methodology even for static networks which utilize the hypersphere geometry.

The remainder of the paper is as follows.  Section \ref{Models} gives the model and methodology.  Section \ref{Estimation} gives estimation methods.  Section \ref{SimulationStudy} describes a simulation study.  Section \ref{DataAnalysis} reports the results from analyzing Newcomb's fraternity data \citep{newcomb1956prediction} and world trade data.  Section \ref{Discussion} gives a discussion.

\section{Models}
\label{Models}
The data we will analyze are of the form $({\cal N},\{{\cal E}_t: t\in \{1,2,\ldots,T\})$, where ${\cal N}$ is the set of all actors (also called by some authors nodes or vertices), and ${\cal E}_t\subseteq \{\{i,j\}, i,j\in {\cal N}, i\neq j\}$ is the set of edges at time $t$.  The edges ${\cal E}_t$ can be viewed as an adjacency matrix $Y_t$ with entries $y_{ijt}$ denoting the edge from actor $i$ to actor $j$ at time $t$.   The latent space approach to modeling networks assumes that there is, for each actor at each time point, a latent position within a network space which represents unobserved actor attributes.  We will assume that at each time point, each actor belongs to one of a fixed number $G$ of clusters; this cluster assignment may change over time.  We will denote the latent position of actor $i$ at time $t$ as ${\bf X}_{it}$ and the cluster assignment for actor $i$ at time $t$ as $\bz_{it}$, a $G$ dimensional vector in which one element is 1 and the others are zero.  We will also let $\cx_t=(\bx_{1t}',\ldots,\bx_{nt}')'$ and $\cz_t=(\bz_{1t}',\ldots,\bz_{nt}')'$.  While the dependency structure of the model may vary, we assume throughout the paper that given the latent positions $\cx_t$, $Y_t$ and $Y_s$, $s\neq t$, are conditionally independent; in many cases (such as binary networks) this assumption can be further extended such that $y_{ijt}$ and $y_{i'j't}$ are conditionally independent given $\cx_t$.

In community detection within a latent space approach, we use the decomposition
\begin{equation}
\pi(\{Y_t,\cx_t,\cz_t\}_{t=1}^T)=\pi(\{Y_t\}_{t=1}^T|\{\cx_t\}_{t=1}^T)\pi(\{\cx_t,\cz_t\}_{t=1}^T).
\end{equation}
The idea here is that the edge probabilities are determined by some underlying attributes which are captured in the latent variables.  Thus if we detect a community in the network it is because there is a corresponding cluster of attributes.  For example, if we see in a social network a group of close friends, this close group, or community, exists because these friends are similar in some fundamental ways, i.e., they have attributes that are clustered together.

\subsection{Distance model}
\label{DistanceModel}
Within the context of the distance model, the network is embedded within a latent Euclidean space, where the probability of edge formation increases as the Euclidean distance between actors decreases.  Let $D({\cx}_t)$ denote the $n\times n$ distance matrix constructed such that $\big(D({\cx}_t)\big)_{ij} \triangleq d_{ijt} = \|\bx_{it}-\bx_{jt}\|$.  In general we will assume that the density of $Y_t$ can be written as a function of the distance matrix $D(\cx_t)$ and some set of likelihood parameters, which we will denote as $\theta_{\ell}$.  For example, the original likelihood for binary networks in \cite{hoff2002latent} is
\begin{equation}
\Prob(y_{ijt}=1|\cx_t,\theta_{\ell})=\frac{\exp\{y_{ijt}\eta_{ijt}\}}{1+\exp\{\eta_{ijt}\}},
\hspace{2pc} \eta_{ijt}=\alpha-d_{ijt}, \label{distLik}
\end{equation}
where in this context $\theta_{\ell}=\{\alpha\}$.
Variants of this likelihood have been proposed, such as in \cite{sarkar2005dynamic}, \cite{krivitsky2009representing}, and \cite{sewell2014latent}.  This last was then extended to account for a wide range of weighted networks in \cite{sewell2014weighted}.  Other likelihoods may be better suited for various other types of weighted edges \citep[see, e.g.,][]{sewell2014analysis}.

\cite{handcock2007model} clustered static network data by clustering the latent positions via a normal mixture model.  This cannot be directly applied to dynamic network data since the latent positions must have some sort of temporal dependency imposed.  Therefore we propose applying the model-based longitudinal clustering model given by \cite{sewell2014model} to the latent positions.  Our focus here is the modeling of the latent positions, which can then be used for whatever likelihood formulation is most appropriate to the data.  We will now describe this model for the latent variables.

We make two assumptions on the latent positions and the cluster assignments.  First, the cluster assignments are assumed to follow a Markov process, i.e.,
$$
\bz_{it}|\bz_{i1},\ldots,\bz_{i(t-1)} \stackrel{{\cal D}}{=}\bz_{it}|\bz_{i(t-1)}.
$$
Second, given the current cluster assignment and all previous cluster assignments and latent positions, we assume the current latent positions depend only on the previous latent positions and the current cluster assignments, i.e.,
$$
\bx_{it}|\bx_{i1},\ldots,\bx_{i(t-1)},\bz_{i1},\ldots,\bz_{it} \stackrel{ {\cal D}}{=}\bx_{it}|\bx_{i(t-1)},\bz_{it}.
$$

The joint density of the latent positions and the cluster assignments is given as
\begin{align}\nonumber
&\pi(\{\cx_t\}_{t=1}^T,\{\cz_t\}_{t=1}^T)&\\
&=\prod_{i=1}^n\prod_{g=1}^G\left[
\beta_{0g}N(\bx_{i1}|\bmu_g,\Sigma_g)
\right]^{Z_{i1g}}
\prod_{t=2}^T
\prod_{h=1}^G\left[
\prod_{k=1}^G\left[
\beta_{hk}N(\bx_{it}|\lambda\bmu_k+(1-\lambda)\bx_{i(t-1)},\Sigma_k)
\right]^{Z_{itk}}
\right]^{Z_{i(t-1)h}},&
 \label{jointXZDist}
\end{align}
where $N(\bx|\bmu,\Sigma)$ is the normal density with mean vector $\bmu$ and covariance matrix $\Sigma$ evaluated at $\bx$.  Thus the communities are each modeled as a multivariate normal distribution in the latent space with mean $\bmu_g$ and covariance matrix $\Sigma_g$.  Since these refer to the location and shape of the $g^{th}$ community in the latent network space, we will refer to $\bmu_g$ and $\Sigma_g$ as the $g^{th}$ community location and community shape respectively.  The mean of the latent position $\bx_{it}$ is then modeled as $\lambda\bmu_g+(1-\lambda)\bx_{i(t-1)}$, $\lambda\in(0,1)$, which is a blending of the current cluster effect $\bmu_g$ with the individual temporal effect $\bx_{i(t-1)}$.  Hence we will refer to $\lambda$ as the blending coefficient.  The $\beta_{0g}$'s determine the probability of initially belonging to the $g^{th}$ community and the $\beta_{hk}$'s determine the probability of transitioning from the $h^{th}$ community to the $k^{th}$ community.  We will therefore refer to the vectors $\bbeta_0=(\beta_{01},\ldots,\beta_{0G})$ and $\bbeta_{h}=(\beta_{h1},\ldots,\beta_{hG})$, $h=1,\ldots,G$, respectively as the initial clustering parameter and the transition parameter for group $h$.

\subsection{Projection model}
\label{ProjectionModel}
\cite{cox1991multidimensional} and \cite{banerjee2005clustering} gave many contexts in which there has been empirical evidence that embedding data onto a hypersphere and/or using cosine distances is preferable to Euclidean space/distances.  Here we continue this tradition by embedding dynamic network data onto the hypersphere.  In this section we assume the more specific, but most commonly encountered, context of directed binary edges (the model to be proposed can be simplified for undirected edges).  In the projection model, every actor is embedded within some latent hypersphere; the probability of an edge forming between two actors depends on the angle, rather than the Euclidean distance, between them.  Thus it is the angle between any two actors that represents the ``closeness" of the actors.  Though the latent space is strictly a Euclidean space rather than a hypersphere, it is more helpful to think of the positions within $\Re^p$ as unit vectors on a $p-1$ dimensional hypersphere with individual edge propensities reflected in the magnitude of the latent positions.

Our proposed likelihood of the adjacency matrices adapts the likelihood of the projection model originally proposed by \cite{hoff2002latent}, and extends \cite{durante2014nonparametric} to allow for directed edges.  The specific form of the likelihood is given as
\begin{align}
\pi(\{Y_t\}_{t=1}^T|\{\cx_t\}_{t=1}^T,\theta_{\ell})&=\prod_{t=1}^T\prod_{i\neq j} \frac{\exp\{y_{ijt}\eta_{ijt}\}}{1+\exp\{\eta_{ijt}\}},&\label{projLik1}\\
\eta_{ijt}&=\alpha+s_j\bx_{it}'\bx_{jt}& \label{projLik2}\\
&=\alpha +\|\bx_{it}\|\cdot(s_j\|\bx_{jt}\|)\cdot\cos(\phi_{ijt}),&\label{projLikAlt}
\end{align}
where $\phi_{ijt}$ is the angle between $\bx_{it}$ and $\bx_{jt}$; in this context $\theta_{\ell}=\{\alpha,\bs\}$, where $\alpha$ reflects a baseline edge propagation rate and $\bs=(s_1,\ldots,s_n)$ is a vector of actor specific parameters that reflect how the tendency of the actors to receive edges relates to the tendency to send edges.  We therefore refer to $\bs$ as the receiver \underline{s}caling parameters.  While (\ref{projLik2}) is simpler, (\ref{projLikAlt}) makes it clear how the probability of an edge from $i$ to $j$ is made up of some constant plus the product of the sending effect of $i$, the receiving effect of $j$, and the closeness between $i$ and $j$ in the latent space as measured by the cosine of the angle between the two actors.

The question remains as to how to perform clustering.  With the projection model the latent positions are embedded within a hypersphere, and thus the clustering must be done in a fundamentally different way than that done for the distance model.  Since we would expect a group of highly connected actors to have small angles between them all, we propose clustering based on the angles of the actors' latent positions.

We first assume that the latent positions follow a hidden Markov model, with the cluster assignments as the hidden states.  That is, the cluster assignments follow a Markov process (i.e., given $\bz_{i(t-1)}$, $\bz_{it}$ is conditionally independent of $\bz_{i(t-s)}$ for any $s>1$), and given the cluster assignments $\cz_t$, the latent positions $\bx_t$ are assumed to be conditionally independent of $\bx_s$ for any $s\neq t$.  

The joint density on the latent positions and cluster assignments is given as
\begin{align}\nonumber
&\pi(\{\cx_t\}_{t=1}^T,\{\cz_t\}_{t=1}^T)&\\
&=\prod_{i=1}^n\prod_{g=1}^G\left[
\beta_{0g}N(\bx_{i1}|r_i\bu_g,\tau_i^{-1}I_p)
\right]^{Z_{i1g}}
\prod_{t=2}^T
\prod_{h=1}^G\left[
\prod_{k=1}^G\left[
\beta_{hk}N(\bx_{it}|r_i\bu_k,\tau_i^{-1}I_p)
\right]^{Z_{itk}}
\right]^{Z_{i(t-1)h}},&
 \label{jointXZProj}
\end{align}
where $I_p$ is the $p\times p$ identity matrix.  As with the distance model of Section \ref{DistanceModel}, the communities are modeled as multivariate normal distributions within the latent space.  Here $\br=(r_1,\ldots,r_n)$, the \underline{r}adii of the means of the $\bx_{it}$'s, are individual effects representing the individual propensities to send edges; hence we refer to $\br$ as the sender propensities.  $\bu_g$ is the unit vector corresponding to the direction of the $g^{th}$ community, and hence we refer to the $\bu_g$'s as the community directions.  $\btau=(\tau_1,\ldots,\tau_n)$ are the precision parameters, and $\bbeta_0=(\beta_{01},\ldots,\beta_{0G})$ and $\bbeta_{h}=(\beta_{h1},\ldots,\beta_{hG})$, $h=1,\ldots,G$, are again respectively the initial clustering parameter and the transition parameter for group $h$.

From (\ref{jointXZProj}) we can see how the different aspects of the network are captured in the joint density of $\{\cx_t\}_{t=1}^T$ and $\{\cz_t\}_{t=1}^T$.  The clusters are completely determined by the community directions $\bu_g$.  Thus if two actors belong to the same cluster then they have the same mean direction, and therefore the model will deem these two actors as similar (based on the cosine of their angle).  The permanence and transience of the clusters are captured in the transition parameters $\bbeta_h$, $h=1,\ldots,G$.  The individual effects are captured by the sender propensities $\br$ and the receiver scaling parameters $\bs$.  To see this more clearly, notice that the square of the individual sending effect (and the scaled individual receiving effect), $\|\bx_{it}\|^2$, has mean $p\tau_i^{-1}+r_i^2$; under the quite reasonable assumption that $\uparrow r_i \notimplies \downarrow \tau_i^{-1}$ (we would expect the opposite to occur), we see that $r_i$ has a direct effect on the individual effect.  The difference in individual $i$'s sending and receiving effect is given by the $i^{th}$ receiver scaling parameter $s_i$.

Note that the parameterization (\ref{projLik2}) of the likelihood (\ref{projLik1}) is not identifiable, as $\bs$ and $\cx_t$ can be scaled arbitrarily.  The estimation is done within a Bayesian framework, however, and thus by fixing the hyperparameters corresponding to the priors on the unknown parameters, the posterior distribution is identifiable.

\section{Estimation}
\label{Estimation}

Our estimation is done within the Bayesian framework, with the goal of finding the maximum {\it a posteriori} (MAP) estimators of the unknown parameters and latent positions.

\subsection{MCMC for the distance model}
\label{MCMC}
We propose a Markov chain Monte Carlo (MCMC) method to obtain posterior modes to estimate the latent positions and model parameters of the distance model given in Section \ref{DistanceModel}.  Specifically, we implement a Metropolis-Hastings (MH) within Gibbs sampler.

We assign the following priors:
\begin{eqnarray}
\lambda&\sim& N_{(0,1)}(\nu_{\lambda},\xi_{\lambda}),\\
\bmu_g&\sim& N({\bf 0},\tau^2I_p) \hspace{1pc}\mbox{for $g=1,\ldots,G$},\\
\Sigma_g&\sim& W^{-1}(p+1,diag(\gamma_1,\ldots,\gamma_p)) \hspace{1pc}\mbox{for $g=1,\ldots,G$},\\
\tau^2&\sim&\Gamma^{-1}(a,b),\\
\gamma_{\ell}&\sim&\Gamma(c,1/d) \hspace{1pc}\mbox{for $\ell=1,\ldots,p$},\\
\boldsymbol\beta_h&\sim&Dir(1,\ldots,1) \hspace{1pc}\mbox{for $h=0,1,\ldots,G$},
\end{eqnarray}
where $N_{(0,1)}(\mu,\sigma^2)$ indicates the normal distribution with mean $\mu$ and variance $\sigma^2$ truncated to the range of $(0,1)$, $W^{-1}(a,B)$ indicates the inverse Wishart distribution with degrees of freedom $a$ and scale matrix $B$, $diag(d_1,\ldots,d_K)$ indicates a $K\times K$ diagonal matrix with $d_1,\ldots, d_K$ on the diagonal, $Dir(a_1,\ldots,a_K)$ indicates the Dirichlet distribution with parameters $a_1$ to $a_K$, $\Gamma^{-1}(a,b)$ indicates the inverse gamma distribution with shape and scale parameters $a$ and $b$ respectively, and $\Gamma(a,b)$ indicates the gamma distribution with shape and scale parameters $a$ and $b$ respectively.  Additionally, there will be some prior $\pi(\theta_{\ell})$ on the likelihood parameters $\theta_{\ell}$ that will depend on the formulation of the likelihood.

With the exception of the latent positions and $\theta_{\ell}$, these priors are conjugate with respect to the full conditional distributions; these distributions are given in the supplementary material.  For the latent positions, MH steps are necessary.  The context specific form of the likelihood will determine whether the likelihood parameters $\theta_{\ell}$ can be sampled directly or whether the user needs to implement MH steps here as well (see Sections \ref{methodEvaluation} and \ref{Fraternity} for examples).

\subsection{Variational Bayesian inference for the projection model}
\label{VB}
\cite{polson2013bayesian} gave a data augmentation scheme for logistic models by utilizing the P\'olya-Gamma distribution.  This scheme starts by introducing a random variable $\omega_{ijt}$ which, given $\eta_{ijt}$, follows $PG(1,\eta_{ijt})$, where $PG(b,c)$ denotes the P\'olya-Gamma distribution with parameters $b>0$ and $c\in\Re$.  This auxiliary variable $\omega_{ijt}$ is conditionally independent of $y_{ijt}$ given $\eta_{ijt}$.  Polson et al. show that the conditional joint density of $y_{ijt}$ and $\omega_{ijt}$ can be written as
\begin{equation}
\pi(y_{ijt},\omega_{ijt}|\eta_{ijt})=
\frac12e^{(y_{ijt}-1/2)\eta_{ijt}}
e^{-\omega_{ijt}\eta_{ijt}^2/2}PG(\omega_{ijt}|1,0),
\label{PGJoint}
\end{equation}
where $PG(\omega|b,c)$ is the P\'olya-Gamma density with parameters $b$ and $c$ evaluated at $\omega$.  This data augmentation leads to tractable forms for the full conditional distributions of the model parameters and latent positions, leading to efficient and accurate estimation for binary data using Gibbs sampling \citep{choi2013polya}, the EM algorithm \citep{scott2013expectation} and, as we will show here, variational Bayes (VB) approaches.

Using Polson et al.'s work we may either implement a Gibbs sampler, as each full conditional distribution belongs to a well known family from which we can sample, or alternatively we may implement a mean field VB algorithm.  Unlike a MCMC approach which obtains samples approximately from the posterior distribution, the VB algorithm here iteratively finds an approximation to the posterior density $\pi(\{\cx_t,\cz_t\}_{t=1}^T,\theta_{\ell},\theta_{p}|\{Y_t\}_{t=1}^T)$, where $\theta_{p}$ is all the remaining model parameters corresponding to the prior on $\{\cx_t,\cz_t\}_{t=1}^T$.  Using the mean field VB implies that we are finding a factorized approximation $Q$ of the posterior which minimizes the Kullback-Liebler divergence between the true posterior and $Q$.  This factorized form will be given shortly.

VB procedures have been gaining popularity in large part due to their greatly decreased computational cost in comparison with most sampling methods.  \cite{salter2013variational} applied VB to the static latent space cluster model for networks given by \cite{handcock2007model} (which is a static form of the distance model).  Within this iterative scheme, the factorized distributions of the latent positions and many of the model parameters required numerical optimization techniques, as a closed form analytical solution was unavailable.  By utilizing the projection model as described in Section \ref{ProjectionModel}, however, we can find closed form solutions for each iteration, thereby reducing the computational cost involved in the estimation algorithm.

We assign the following priors:
\begin{align}
\omega_{ijt}&\sim PG(1,0) \hspace{1pc}\mbox{for $t=1,\ldots,T$, $1\leq i \neq j \leq n$},&\\
s_i&\sim Exp(1)\hspace{1pc}\mbox{for $i=1,\ldots,n$},&\\
r_i|\tau_i&\sim \Gamma(1,c\tau_i^{-1})
\hspace{1pc}\mbox{for $i=1,\ldots,n$},&\\
\tau_i&\sim \Gamma(a_2^*,b_2^*)
\hspace{1pc}\mbox{for $i=1,\ldots,n$},&\\
\alpha&\sim N(0,b_3^*), &\\
\pi(\bu_g)&=\frac{\Gamma(p/2)}{2\pi^{p/2}}\hspace{1pc}\mbox{for $h=0,1,\ldots,G$},&\\
\bbeta_h&\sim Dir(\boldsymbol\gamma_h^*) \hspace{1pc}\mbox{for $h=0,1,\ldots,G$}.&
\end{align}
To estimate the posterior $\pi(\{\cx_t,\cz_t\}_{t=1}^T,\theta_{\ell},\theta_p|\{Y_t\}_{t=1}^T)$, we use the factorized approximation $Q$, which looks like
\begin{align}\nonumber
&Q(\Omega,\{\cx_t\}_{t=1}^T,\{\cz_t\}_{t=1}^T,\alpha,\bs,\br,\btau,\bu,\{\bbeta_h\}_{h=0}^G)&\\
&=q(\Omega)q(\{\cx_t\}_{t=1}^T)q(\{\cz_t\}_{t=1}^T)q(\alpha)q(\bs)q(\br)q(\btau)q(\bu)q(\{\bbeta_h\}_{h=0}^G),&
\label{factorized}
\end{align}
where $\Omega=\{\omega_{ijt}\}_{t,i\neq j}$.  Using the priors given above, the factorized distributions on the right hand side of (\ref{factorized}) all belong to well known families of distributions.  The exact forms are given in the supplementary material.

\begin{figure}
\centering
\includegraphics[height=7cm]{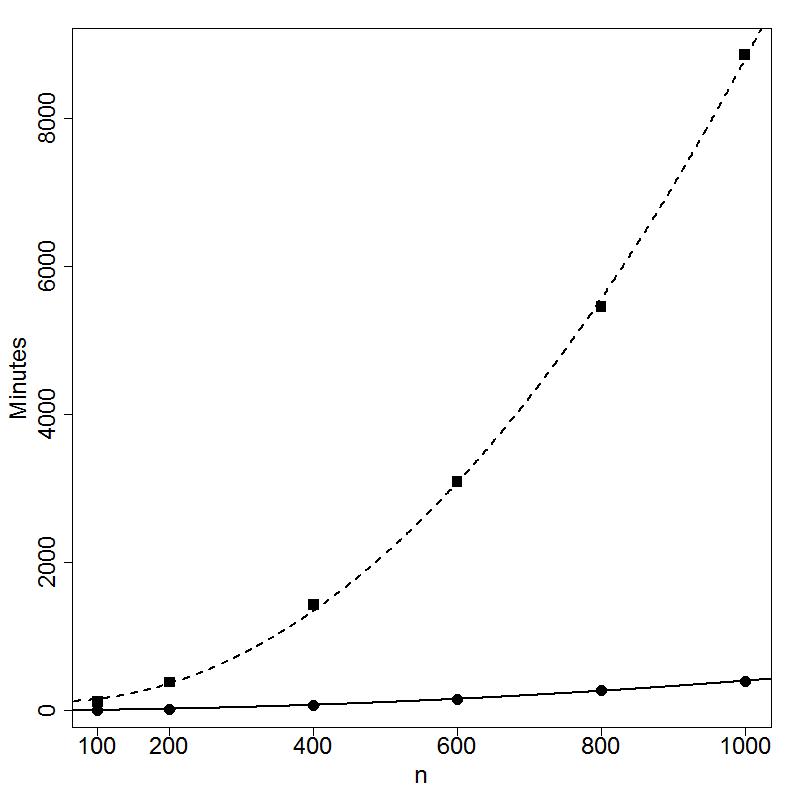}
\caption{Run time in minutes for 50,000 draws using the MCMC algorithm (dashed line, squares) and 500 iterations of the VB algorithm (solid line, circles)}
\label{compTime}
\end{figure}

Of interest is the computational time required for our proposed methods, and in particular how the VB algorithm decreases the computational time required.  We recorded the times required to implement both our VB approach (500 iterations) and the corresponding Gibbs sampler (50,000 samples drawn), letting $n$ be 100, 200, 400, 600, 800, and 1,000.  The times are given graphically in Figure \ref{compTime}.  From this we can see that the VB algorithm shows drastic reduction in computational cost.  We will see in Section \ref{SimulationStudy}, however, that the performance of the VB and Gibbs sampler are very similar.

\subsection{Initialization}
Our context involves a high dimensional estimation problem, and so how we initialize the MCMC or the VB algorithm plays a non-negligible role in the performance.  We performed a small sensitivity analysis of the starting conditions of our algorithms, the details of which can be found in the supplementary material. The results indicated that under some conditions the VB algorithm for the projection model can be sensitive to the initialization scheme, though it did not appear that either of the MCMC algorithms (the Gibbs sampler for the projection model and the MH within Gibbs sampler for the distance model) were particularly sensitive.  The full details on how we initialized the algorithms are given in the supplementary material.

\subsection{Number of communities}
\label{NumberOfClusters}
An implicit challenge underlying the previous discourse is that in practice we do not in general know the number of communities $G$.  We found the strategy given by \cite{handcock2007model} to be quite successful in our simulation study (see Section \ref{BICsimstudy}).  We briefly summarize this method and refer the interested reader to the original source for more details.

Rather than estimating the integrated likelihood $\pi(\{Y_t\}_{t=1}^T|G)$ as would typically be done, we instead consider the joint distribution of the observed network data and unobserved latent positions, using our MAP estimator as the fixed values of the latent positions, i.e., $\pi(\{Y_t\}_{t=1}^T,\{\widehat{\cx}_t\}_{t=1}^T|G)$, where $\{\widehat{\cx}_t\}_{t=1}^T$ is the MAP estimators of the latent positions.  We can rewrite this as
\begin{equation}
\pi(\{Y_t\}_{t=1}^T,\{\widehat{\cx}_t\}_{t=1}^T|G)=\int \pi(\{Y_t\}_{t=1}^T|\{\widehat{\cx}_t\}_{t=1}^T,\theta_{\ell})\pi(\theta_{\ell})d\theta_{\ell} \int \pi(\{\widehat{\cx}_t\}_{t=1}^T|\theta_p)\pi(\theta_p)d\theta_p,
\label{intLik}
\end{equation}
where all distributions are implicitly conditioning on $G$.  The two integrals on the right hand side of (\ref{intLik}) can each be estimated via the Bayesian information criterion (BIC), thus allowing us to find the BIC approximation of $2\log(\pi(\{Y_t\}_{t=1}^T,\{\widehat{\cx}_t\}_{t=1}^T|G))$ as
\begin{equation*}
\mbox{BIC}=\mbox{BIC}_1+\mbox{BIC}_2,
\end{equation*}
where
\begin{align} \nonumber
\mbox{BIC}_1&=2\log(\pi(\{Y_t\}_{t=1}^T|\{\widehat{\cx}_t\}_{t=1}^T,\hat\theta_{\ell}))  - \mbox{dim}(\theta_{\ell})\log\Big(\sum_{t,i\neq j} y_{ijt}\Big), &\\ \nonumber
\mbox{BIC}_2&= 2\log(\pi(\{\widehat{\cx}_t\}_{t=1}^T|\hat\theta_p)) -  \mbox{dim}(\theta_p)\log(nT).&
\end{align}

Rather than using maximum likelihood estimators for $\hat\theta_{\ell}$ and $\hat\theta_{p}$ in computing the BIC's, we used the MAP estimators, as was also done in, e.g., \cite{fraley2007bayesian}.  We remark that for the projection model, since the posterior modes found by the VB and the Gibbs sampler perform comparably (see Section \ref{methodEvaluation}), this BIC model selection method is still valid for the VB estimates.  This is because we only need the posterior mode, and hence any inaccuracies in the posterior variances/covariances of the parameters induced by approximating the posterior distribution with the VB factorized distribution will not affect the BIC criterion.  One last note is that we utilized recursive relations identical or similar to those given in \cite{sewell2014model} in order for the number of terms required to compute $\pi(\{\cx_t\}_{t=1}^T|\hat\theta_p)$ to be linear, rather than exponential, with respect to $T$.

\section{Simulation study}
\label{SimulationStudy}
\subsection{Method evaluation}
\label{methodEvaluation}
We simulated 200 binary networks, each with $n=100$ actors and $T=10$ time points.  These 200 data sets were subdivided evenly in two different ways.  First, half of the data sets were generated according to the distance model, the other half via the projection model.  Second, half of the data sets had sticky cluster transition probabilities, letting the $\beta_{hh}$'s take large values (recall that $\bbeta_h$ is the transition parameter for group $h$), while the other half had more transitory transition probabilities, letting the $\beta_{hh}$'s to take more moderate values.  In summary, we had 50 data sets from the distance model with sticky transition probabilities, 50 from the distance model with transitory transitions, 50 from the projection model with sticky transition probabilities, and 50 from the projection model with transitory transitions.  Details on how the data were generated will be given shortly.

We compared various methods in four ways.  The first was to evaluate how well the model explains the data used to fit the model.  To this end we obtained in-sample edge predictions and computed the AUC  (area under the receiver operating characteristic curve);  a value of one implies a perfect fit, whereas a value of 0.5 implies that the predictions are random.  As a good in-sample fit may be due to overfitting the data, we also looked at one step ahead predictions.  We obtained one step ahead predicted probabilities and computed the correlation with the true one step ahead probabilities.  We aim to stress, however, that prediction is not the primary purpose of this methodology, but rather to accurately recover hidden communities in the network object.  We thus compared the true clustering assignments with the estimated clustering assignments using two methods.  The first is the corrected Rand index (CRI), which can be viewed as a measure of misclassification.  Values close to 1 indicate nearly identical clustering assignments and values near zero indicate what one might expect with two random clustering assignments.  Second, we computed the variation of information (VI) \citep{meilua2003comparing}.  The VI is a true metric, and hence a smaller VI value implies that the two clusterings being compared are closer to being identical.

For each of the 200 simulations we compared six methods.  The first two are the VB algorithm and the Gibbs sampler for the projection model.  The third is the distance model.  Here we used the likelihood formulation found in the dynamic latent space model of \cite{sewell2014latent}.  This likelihood is given as
\begin{equation}
\logit(\Prob(y_{ijt}=1|\cx_t,\beta_{IN},\beta_{OUT},s_i,s_j))=\beta_{IN}\Big(1-\frac{d_{ijt}}{s_j}\Big)+\beta_{OUT}\Big(1-\frac{d_{ijt}}{s_i}\Big),
\label{SandC2014Likelihood}
\end{equation}
where $\beta_{IN}$ and $\beta_{OUT}$ are global parameters that reflect the relative importance of popularity and activity respectively, the $s_i$'s are actor specific parameters that reflect the tendency to send and receive edges, and $d_{ijt}$ is the distance between actors $i$ and $j$ within the latent Euclidean space at time $t$.  Estimation is done by putting a bivariate normal prior on $\beta_{IN}$ and $\beta_{OUT}$, a Dirichlet prior on the $s_i$'s, and incorporating these parameters in the MH within Gibbs MCMC algorithm of Section \ref{MCMC}.  The fourth and fifth methods were the clustering models of \cite{handcock2007model} and of \cite{krivitsky2009representing}, implemented in the latentnet R package \citep{krivitsky2008fitting,krivitsky2015latentnet}.  These latter two models cluster static networks via a latent space approach; to apply them to dynamic networks, clustering was performed at each time point and then combined sequentially using the relabeling algorithm given in \cite{papastamoulis2010artificial}.  Note that these two methods, being static models, cannot be used to perform one step ahead predictions.  Lastly we used the temporal exponential random graph model (TERGM) \citep{hanneke2010discrete}, as implemented in the btergm R package \citep{leifeld2015xergm}.  The terms we specified for the TERGM were the total number of edges in the network, the number of reciprocated edges, the number of transitive triples, the number of cyclic triples, in-degrees and out-degrees, the number of lagged reciprocated edges, and the stability of the network.  Note that this method can be used to determine in-sample predictions and one step ahead predictions, but has no functionality for determining cluster assignments.  All MCMC methods were used to obtain 50,000 samples.

For the data sets generated according to the distance model, we set the blending coefficient $\lambda=0.8$, the dimension of the latent space $p=2$, the total number of clusters $G=6$, and the likelihood parameters $\beta_{IN}=0.3$, and $\beta_{OUT}=0.7$.  We set the community locations $\bmu_g$ to be $(-0.03,0)$, $(-0.01,0)$, $(0.01,0)$, $(0.03,0)$, $(0,0.02)$, and $(0,-0.02)$.  We drew the community shapes $\Sigma_g$ , $g=1,\ldots,G$, from $W^{-1}(13,(1\times10^{-5}) I_2)$, the initial clustering parameter $\bbeta_0\sim Dir(10,\ldots,10)$, and for $h=1,\ldots,6$, the transition parameter $\bbeta_h$  for group $h$ was set to was set to be proportional to
\begin{equation}\nonumber
\left(\frac{1}{\|\boldsymbol\mu_1 -\boldsymbol\mu_h\|},\ldots,\frac{1}{\|\boldsymbol\mu_{h-1} -\boldsymbol\mu_h\|},\mbox{const}\times \max_{k\neq h}\left\{\frac{1}{\|\boldsymbol\mu_k -\boldsymbol\mu_h\|}\right\},\frac{1}{\|\boldsymbol\mu_{h+1} -\boldsymbol\mu_h\|},\ldots,\frac{1}{\|\boldsymbol\mu_K -\boldsymbol\mu_h\|} \right).
\end{equation}
For sticky transition probabilities we set the constant in the above equation equal to 20 which yields probabilities  from 0.82 to 0.87 of remaining in the same cluster, and for transitory transition probabilities we set the constant equal to 10 which yields probabilities from 0.70 to 0.77 of remaining in the same cluster.
The cluster assignments $\{\cz_t\}_{t=1}^T$ and latent positions $\{\cx_t\}_{t=1}^T$ were drawn according to (\ref{jointXZDist}), and the actor specific parameters $(s_1,\ldots,s_n)\sim Dir\Big(100\frac{1/\|X_{1,1}\|}{\max_j(1/\|X_{j,1}\|)},\ldots,100\frac{1/\|X_{100,1}\|}{\max_j(1/\|X_{j,1}\|)}\Big)$.  Finally, the adjacency matrices were simulated according to (\ref{SandC2014Likelihood}).  This led to an average density of the simulated networks (taken over all time points of all simulations) of 0.221 and 0.222 for sticky and transitory transition probabilities respectfully.  The average modularity (again averaged over all time points of all simulations) was 0.299 and 0.287 for sticky and transitory transition probabilities respectfully, giving a measure of how well separated the clusters are.  Specifically, the modularity  \citep[originally defined by][for undirected networks]{clauset2004finding} as implemented in the igraph package \citep{igraphPackage} is
$$
\frac{1}{2S_t}\sum_{i\neq j}\left(Y^*_{ijt}-\frac{k_{it}k_{jt}}{2S_t}\right)1_{\{ \bz_{it}'\bz_{jt}=1 \}},
$$
where $S_t$ is the number of edges in the network at time $t$, $Y_t^*$ is the $n\times n$ symmetric adjacency matrix constructed by setting $Y_{ijt}^*=Y_{jit}^*=(Y_{ijt}+Y_{jit})/2$, and $k_{it}$ is the average of the in degree and out degree for actor $i$ at time $t$.  For comparison, an Erd\H{o}s-R\'enyi graph with comparable density has on average a modularity of 0.076, and a network consisting of 5 fully connected subgraphs, each of which are fully disconnected, has a modularity of 0.8 (and a density of 0.19).

For the data sets generated according to the projection model, we set the total number of clusters $G=6$, the dimension of the latent space $p=3$, the baseline propagation rate $\alpha=-5$, the initial clustering parameter $\bbeta_0 = (1/6,\ldots,1/6)$ and the community directions
$$
(\bu_1,\ldots,\bu_6) = \left[\begin{array}{cccccc}
-15&30&60&105&45&45 \\
0   &0&   0&    0&60&-60
\end{array}\right],
$$
where $\bu_g$ are given in the spherical coordinate angles in degrees.  For $h=1,\ldots,6$, the transition parameter $\bbeta_h$ was set to be proportional to $(\exp(\mbox{const}\cdot\bu_h'\bu_1),\ldots,\exp(\mbox{const}\cdot\bu_h'\bu_6))$.  For sticky transition probabilities we set the constant above equal to 8 which yields probabilities from 0.68 to 0.96 of remaining in the same cluster, and for transitory transition probabilities we set the constant equal to 5 which yields probabilities from 0.52 to 0.83 of remaining in the same cluster.
For $i=1,\ldots,100$, we simulated the receiver scaling parameters $s_i\sim N(1,0.15)$,  the sender propensities $r_i\sim N(2.3,0.05^2)$, and set the precision parameters $\tau_i=175/r_i^2$.  The cluster assignments $\{\cz_t\}_{t=1}^T$ and latent positions $\{\cx_t\}_{t=1}^T$ were drawn according to (\ref{jointXZProj}).  Finally, the adjacency matrices were simulated according to (\ref{projLik1}) and (\ref{projLik2}).  This led to an average modularity of 0.305 and 0.279 for sticky and transitory transition probabilities respectfully.  The average density of the simulated networks was 0.183 and 0.191 for sticky and transitory transition probabilities respectfully.

\begin{figure}[t]
\centering
\includegraphics[width=0.7\textwidth]{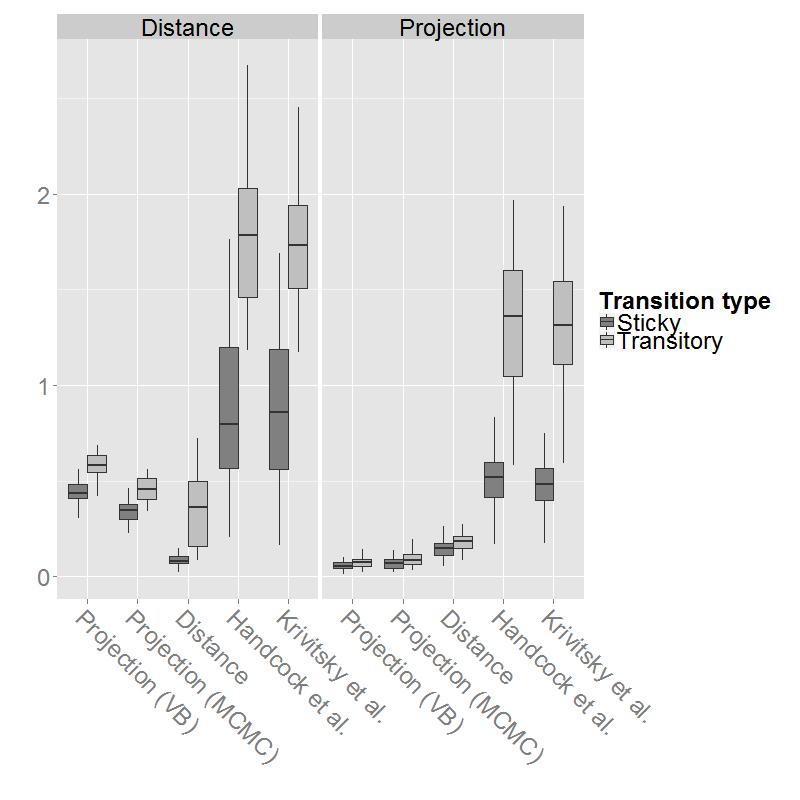}
\caption{The Variation of Information (VI) values from the simulation study are given here graphically, separated by the underlying true geometry (distance/projection) and the type of transition (sticky/transitory).}
\label{simVI}
\end{figure}

Table \ref{simTable} gives the simulation results.  The AUC values show that the TERGM fits the data poorly, but all the other methods fit rather comparably.  However, looking at the CRI and VI we see that the static methods are overfitting the model; that is, they are providing good predicted probabilities for the observed data used to fit the model but are not doing so well at capturing the underlying truth.  The correlation between the estimated one step ahead probabilities and the true probabilities are much higher for our methods than for the TERGM.  Note that both the projection model and the distance model provide good predictive performance regardless of the true geometry of the latent space and regardless of the cluster transition probability matrix.  Once we start looking at the CRI and the VI, which is of primary importance with respect to the goals of the proposed work, we notice several things.  First, when using the projection model, the VB and the Gibbs sampler yield very similar performance.  Second, when the geometry of the latent space is misspecified, our proposed models still perform quite well and in fact perform similarly to the correctly specified model.  Lastly, we note that the performance of the static methods deteriorate when the probabilities of changing clusters increase.  The VI values are also given graphically in Figure \ref{simVI}, visually demonstrating the performance disparities between the dynamic and static methods.
\begin{landscape}
\begin{table}[p]
\centering
\footnotesize
\begin{tabular}{  l ll l l l l l }
\hline
	True model & Transitions & Fitted model & AUC (in sample) & Correlation (one step ahead) & CRI & VI \\ \hline
	Projection & Sticky & Projection (VB) & 0.889 (0.00579) & 0.987 (0.00376) & 0.987 (0.00967) & 0.0560 (0.0284) \\
	Projection & Sticky & Projection(MCMC) & 0.885 (0.00601) & 0.975 (0.00401) & 0.984 (0.00889) & 0.0676 (0.0265)  \  \\
	Projection & Sticky & Distance & 0.875 (0.00623) & 0.933 (0.0155) & 0.954 (0.0182) & 0.150 (0.0491)  \  \\
	Projection & Sticky & Handcock et al. & 0.876 (0.00664)  & NA & 0.799 (0.0893) & 0.518 (0.192)  \  \\
	Projection & Sticky & Krivitsky et al. & 0.899 (0.00543) & NA & 0.806 (0.0866) & 0.485 (0.185)  \  \\
	Projection & Sticky & TERGM & 0.619 (0.0154) & 0.270 (0.114) & NA & NA   \\
	\  & \  & \  & \  & \  & \  & \  & \  \\
	Projection & Transitory& Projection (VB) & 0.884 (0.00444) & 0.980 (0.0130) & 0.981 (0.0767) & 0.0741 (0.193) \\
	Projection & Transitory & Projection(MCMC) & 0.880 (0.00458) & 0.962 (0.0153) & 0.977 (0.0743) & 0.0862 (0.187)   \\
	Projection & Transitory & Distance & 0.870 (0.00458) & 0.922 (0.0222) & 0.944 (0.0692) & 0.183 (0.170) \\
	Projection & Transitory & Handcock et al. & 0.871 (0.00497) & NA & 0.520 (0.109) & 1.36 (0.328)  \\
	Projection & Transitory & Krivitsky et al. & 0.895 (0.00392)  & NA & 0.528 (0.113) & 1.31 (0.335)  \\
	Projection & Transitory & TERGM & 0.618 (0.0144) & 0.261 (0.0749) & NA & NA  \\
	\  & \  & \  & \  & \  & \  & \  & \  \\
	Distance & Sticky& Projection (VB) & 0.862 (0.00523) & 0.858 (0.0287) & 0.876 (0.0404) & 0.436 (0.102) \\
	Distance & Sticky & Projection(MCMC) & 0.855 (0.00585) & 0.863 (0.0275) & 0.903 (0.0428) & 0.349 (0.104)  \\
	Distance & Sticky & Distance & 0.863 (0.00575) & 0.928 (0.0303) & 0.981 (0.0542) & 0.0821 (0.129)  \\
	Distance & Sticky & Handcock et al. & 0.861 (0.00518) & NA & 0.733 (0.141) & 0.798 (0.406)  \\
	Distance & Sticky & Krivitsky et al. & 0.879 (0.00520) & NA & 0.719 (0.1334) & 0.861 (0.391)  \\
	Distance & Sticky & TERGM & 0.601 (0.0146) & 0.293 (0.0663) & NA  & NA  \\
	\  & \  & \  & \  & \  & \  & \  & \  \\
	Distance & Transitory & Projection (VB) & 0.853 (0.00667) & 0.845 (0.0338) & 0.820 (0.0751) & 0.583 (0.202) \\
	Distance & Transitory & Projection(MCMC) & 0.846 (0.00702) & 0.834 (0.0272) & 0.864 (0.0731) & 0.455 (0.197) \\
	Distance & Transitory & Distance & 0.851 (0.00644) & 0.882 (0.0351) & 0.889 (0.122) & 0.364 (0.295)  \\
	Distance & Transitory & Handcock et al. & 0.851 (0.00572) & NA & 0.418 (0.115) & 1.78 (0.371)  \\
	Distance & Transitory & Krivitsky et al. & 0.872 (0.00585)  & NA & 0.421 (0.102) & 1.73 (0.311)  \\
	Distance & Transitory & TERGM & 0.597 (0.0140) & 0.224 (0.0378) & NA & NA  \\
\end{tabular}
\normalsize
\caption{Simulation results from data generated according to the distance and projection models with both sticky and transitory cluster transition probabilities.  The median values are reported, with standard deviations in parentheses.  The AUC corresponds to the data used to fit the model, the Correlation (one step ahead) values correspond to the correlation between the estimated probabilities and the true probabilities, the CRI and VI are the corrected Rand index and variation of information respectively between the true and estimated cluster assignments.}
\label{simTable}
\end{table}
\end{landscape}

\subsection{Sensitivity study}
It is not obvious how to choose the values of the hyperparameters from Section \ref{Estimation}.  In the above simulation study as well as in Section \ref{DataAnalysis}, we used an automatic selection method for these hyperparameters, the details of which can be found in the supplementary material.  It is important, however, to determine how sensitive the estimation procedures are to the choice of hyperparameters.  To this end we analyzed 100 data sets simulated according to the projection model and 100 according to the distance model, in each case fitting the data using the model with the correct geometry.  Each set of 100 data sets was evenly divided between sticky and transitory cluster transition probabilities.  For each simulation we evaluated the clustering performance using CRI and VI.

For each simulation we set the hyperparameters in the following way.  For the distance model, we drew $\nu_\lambda\sim Unif(0.5,1)$ and fixed $\xi_{\lambda}=1$, fixed $a=3$ and drew $b\sim Unif(0.01,0.05)$, fixed $c=1.001$ and drew $d\sim N(10,2.5^2)$.  For the projection model, we drew $c\sim \Gamma(20,0.5)$, $a_2^*\sim N(600,100^2)$, and $b_2^*\sim \Gamma(1,0.05)$, and fixed $b_3^*=100$.

Table \ref{priorSensTab} provides the results from this sensitivity analysis.  From this we see that the projection models still perform quite well, although the Gibbs sampler for the projection model has a larger standard deviation of the performance measures.  What we should immediately notice is the appalling performance of the distance model when $\xi_\lambda=1$.  Upon closer inspection we noticed that the parameter estimates of the blending coefficient $\lambda$ were in nearly all cases very close to zero, which means that the model was not using much of the cluster information to predict the latent positions.  As a remedy, we altered this part of the sensitivity analysis, drawing $\nu_\lambda\sim Unif(0.7,0.95)$ and fixing $\xi_\lambda=5\times10^{-4}$, thereby setting a very low prior probability that $\lambda$ is small.  With this alteration we see from Table \ref{priorSensTab} that the clustering performance is quite satisfactory.  In summary, the estimation methods are not particularly sensitive to the selection of hyperparameters with the exception of those associated with $\lambda$.

\begin{table}[htb]
\centering
\begin{tabular}{  l  l  l  l  }
\hline
	Transitions & Fitted model & CRI & VI \\ \hline
	Sticky & Distance ($\xi_{\lambda}=1$) & 0.0169 (0.0104) & 3.42 (0.0756) \\
	Sticky & Distance ($\xi_{\lambda}=5\times10^{-4}$) & 0.981 (0.0427) & 0.0921 (0.113) \\
	Sticky & Projection (VB) & 0.989 (0.00844) & 0.0477 (0.0263) \\
	Sticky & Projection (MCMC) & 0.976 (0.194) & 0.0904 (0.635) \\
	\  & \  & \  & \  \\
	Transitory & Distance ($\xi_{\lambda}=1$) & 0.0218 (0.0341) & 3.382 (0.129) \\
	Transitory & Distance ($\xi_{\lambda}=5\times10^{-4}$) & 0.917 (0.139) & 0.322 (0.362) \\
	Transitory & Projection (VB) & 0.980 (0.136) & 0.0734 (0.353) \\
	Transitory & Projection (MCMC) & 0.962 (0.219) & 0.126 (0.681) \\
\end{tabular}
\caption{Simulation results testing prior sensitivity for data generated according to the distance and projection models with both sticky and transitory cluster transition probabilities.  The median values are reported, with standard deviations in parentheses.}
\label{priorSensTab}
\end{table}

\subsection{BIC model selection}
\label{BICsimstudy}
The last simulation study evaluates the BIC method described in Section \ref{NumberOfClusters}.  Due to the increased computational cost to fit the model for several values of $G$, we generated 15 data sets each from the distance model and the projection model (30 total).  We fitted both the distance and projection models to each data set for $G\in\{3,\ldots,9\}$, and selected the $G$ with the optimal BIC value.

One important comment is that the BIC method of Section \ref{NumberOfClusters} is not appropriate to select the geometry of the latent space, i.e., choose whether we should use the distance or the projection model.  Instead we used the deviance information criterion (DIC) \citep{spiegelhalter2002bayesian} to make this distinction.  We originally attempted to use DIC to choose both the geometry and the number of clusters, but DIC performed extremely poorly at determining $G$.  DIC was, however, perfect at selecting the geometry (in this simulation study) once the optimal number of clusters had been chosen (via BIC).  Therefore based on this simulation study, we recommend to the practitioner the admittedly inelegant procedure of first using the BIC (as described in Section \ref{NumberOfClusters}) to choose $G$ for each geometry, and then using DIC to compare these two models with differing geometries.

\begin{figure}
\centering
\includegraphics[width=0.6\textwidth]{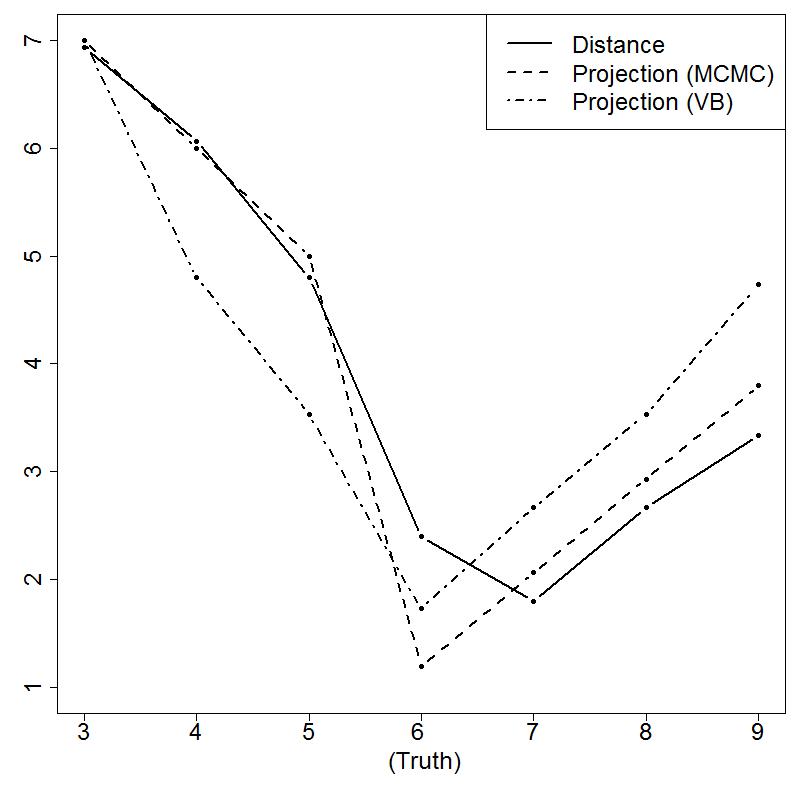}
\caption{Simulation results testing the BIC method of selecting the number of clusters $G$ (horizontal axis).  The vertical axis represents the average rankings over 15 simulations (for each model), where low values indicate better BIC values.  The true number of clusters is 6.}
\label{BICFigure}
\end{figure}

Figure \ref{BICFigure} provides the results.  As mentioned above, DIC perfectly selected the geometry, and so we only present the BIC values for the model with the correctly specified geometry for varying $G$.  Specifically, Figure \ref{BICFigure} gives the average ranking of the BIC values, where low rankings indicate better BIC values.  From this we see that the true number of clusters (6) is frequently chosen as the optimal number of clusters, and values of $G$ far from the truth rank poorly.

\section{Data analysis}
\label{DataAnalysis}
\subsection{Newcomb's fraternity data}
\label{Fraternity}
\cite{newcomb1956prediction} discussed data collected on 17 male college students who were previously unknown to each other.  These 17 students, as part of Newcomb's study, agreed to live together for sixteen weeks (though the data set excludes the ninth week due to school vacation).  For each week, every student ranks the other 16 students from 1 (most favored) to 16 (least favored).  

In this context $Y_t$ is the $t^{th}$ $n\times n$ adjacency matrix whose $i^{th}$ row, denoted ${\bf y}_{it}$, is how the $i^{th}$ actor ranks the other $n-1$ actors.  Without loss of generality, assume that the rankings go, in order of most favored to least favored, from 1 to $n-1$.  Then we let $\boldsymbol{o}_{it}=(o_{i1t},o_{i2t},\ldots,o_{i(n-1)t})$ denote the $(n-1)\times1$ vector which is the ordering of the rank vector ${\bf y}_{it}$ (e.g., if ${\bf y}_{1t}=(0,4,3,1,2)$ then $\boldsymbol{o}_{1t}=(4,5,3,2)$).
We assume that, conditioning on $({\cal X}_t,\boldsymbol\Psi)$, ${\bf y}_{it}$ is independent of ${\bf y}_{i't}$, $i\neq i'$.

The likelihood we will use is that used by \cite{sewell2014analysis}, given as
\begin{equation}
\mathbb{P}(Y_t|{\cal X}_t,\bs) =\prod_{i=1}^n\prod_{j=1}^{n-1}\frac{s_{o_{ijt}}\exp(-d_{io_{ijt}t}) }{\sum_{\ell=j}^{n-1}s_{o_{i\ell t}}\exp(-d_{io_{i\ell t}t})},
\label{PL2}
\end{equation}
where again $\bs=(s_1,\ldots,s_n)$ are actor specific parameters which indicate an actor's social reach, and for identifiability $\sum_{i=1}^ns_i=1$.  This is a Plackett-Luce model \citep{plackett1975analysis}, and as such satisfies Luce's Choice axiom which can be characterized by having actor $i$ rank actor $j$ over actor $k$ with the same probability whether or not actor $\ell$ is included in the set to be ranked.  See \cite{sewell2014analysis} for further motivation and details of this model.  As this likelihood depends on the latent positions through the distances $D(\cx_t)$'s, we implement the distance model of Section \ref{DistanceModel}.  This flexible framework allows us to detect communities through the latent positions of the students.  Estimation is done by putting a Dirichlet prior on $\bs$ and incorporating these parameters in the MH within Gibbs MCMC algorithm of Section \ref{MCMC}.

For $G=2,\ldots,9$, we ran 100,000 iterations of the MCMC algorithm of Section \ref{MCMC}, thus having a maximum of nine clusters.  For each of the 8 chains, we used a short MCMC chain (the same chain for each $G$) following the model with no clustering of \cite{sewell2014analysis} to initialize the latent positions $\{\cx_t\}_{t=1}^T$ and the actor specific likelihood parameters $\bs$, and for the remaining prior parameters we used the generalized EM algorithm given by \cite{sewell2014model}.

\begin{figure}
\centering
\begin{subfigure}[b]{0.4\textwidth}
\includegraphics[height=\textwidth]{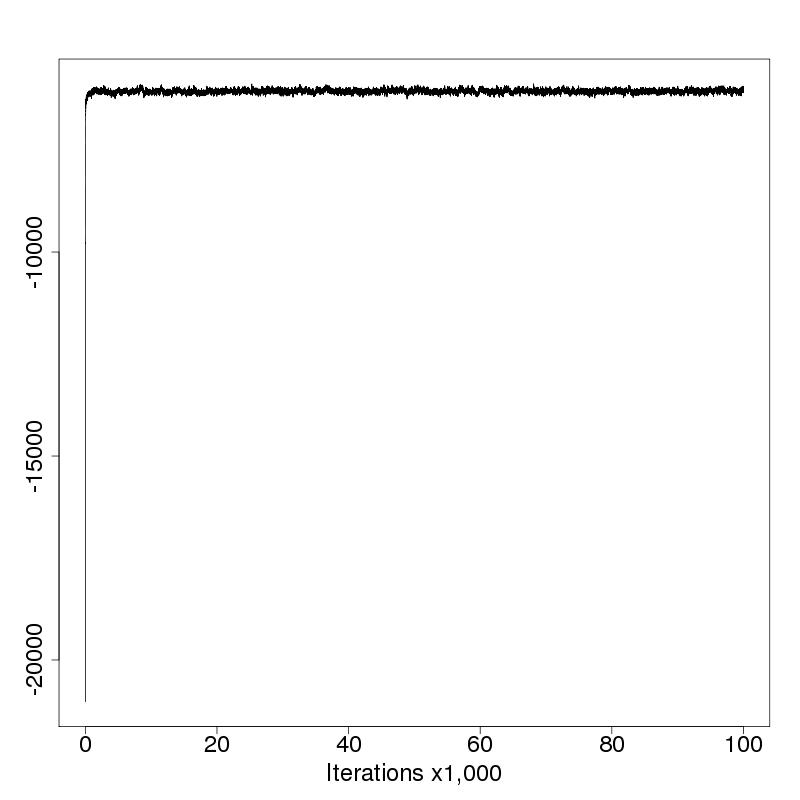}
\caption{Trace plot of the posterior value for each iteration of the MCMC algorithm.}
\label{fratTrace}
\end{subfigure}
\hspace{3pc}
\begin{subfigure}[b]{0.4\textwidth}
\includegraphics[height=\textwidth]{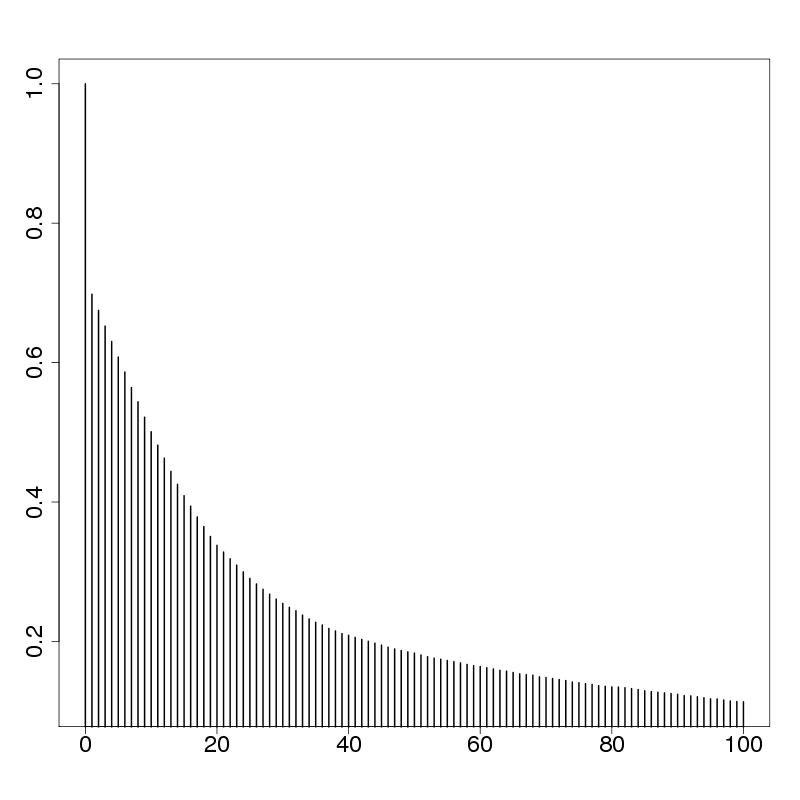}
\caption{Autocorrelation function (ACF) plot.}
\label{fratACF}
\end{subfigure}
\caption{Diagnostic plots for MCMC estimation corresponding to the fraternity data.}
\end{figure}

The BIC method described in Section \ref{NumberOfClusters} led us to choose five communities.  These BIC values ranged from $-13,531$ to $-13,066$.  The MCMC chain converged relatively quickly, as is seen in Figure \ref{fratTrace}, which provides a trace plot of the posterior value for all 100,000 samples.  Adjacent in Figure \ref{fratACF} is the ACF plot, which shows that the correlation decays at a reasonable rate, and, together with Figure \ref{fratTrace} indicates that we had good mixing.  Geweke's diagnostic test, as implemented in the coda R package \citep{codaRpackage}, yielded a $p$-value of 0.611 using a burn in of 5,000, implying convergence.

The goodness of fit was evaluated using the pseudo-$R^2$ value described in \cite{sewell2014analysis}.  The pseudo-$R^2$ takes values in the interval $[0,1)$, where a higher value implies a better fit of the data.  After analyzing the data, we obtained a pseudo-$R^2$ value of 0.575.  This is slightly less than that obtained by Sewell and Chen (0.622), which we feel satisfied with since we are imposing more structure via the clustering on the prior of the latent positions; that is, though we are imposing more structure on the prior of the latent positions, we are not losing much in terms of model fit.

This data set has been analyzed many times since its genesis, and several of these analyses have focused at least in part on community detection.  \cite{nakao1993longitudinal}, when analyzing Newcomb's fraternity data, created similarity matrices for each time point and then performed multidimensional scaling to obtain latent network positions, visually determining the communities.  \cite{moody2005dynamic} used various visualization methods and also commented on some clustering that were noticed via visual inspection.  \cite{sewell2014analysis} provided a detailed analysis of Newcomb's fraternity data which included a post-hoc analysis of the subgroup formation.  

\begin{figure}[t]
\centering
\includegraphics[width=0.75\textwidth]{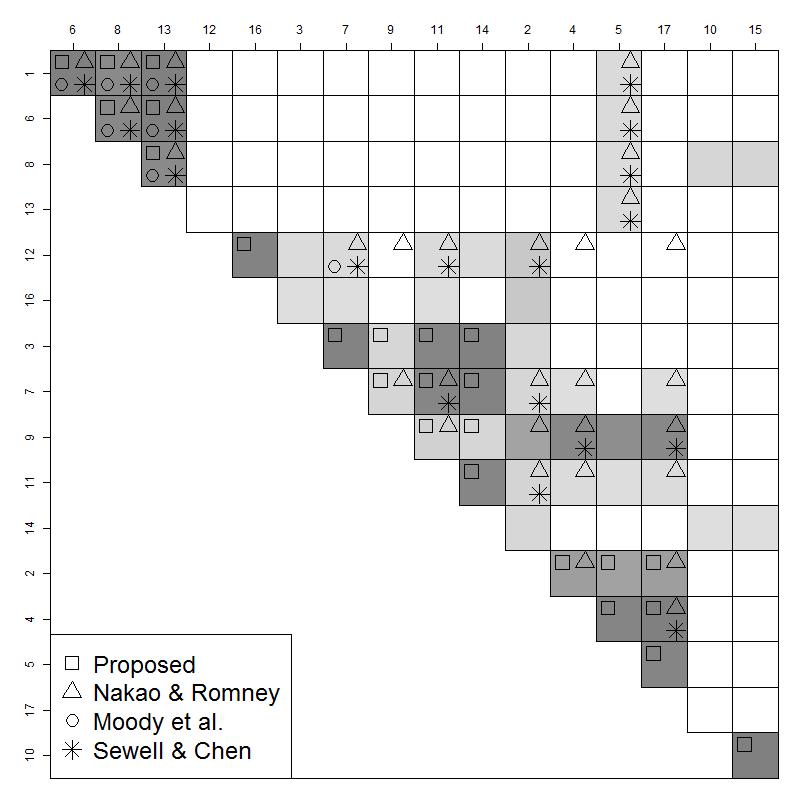}
\caption{Pairwise probabilities of actors belonging to the same cluster at week 7.  Actors (rows/columns) are ordered according to the MAP estimates of the communities.  Different methods' estimates are given by the methods' corresponding shapes in the appropriate cell.  Shown are our proposed MAP estimates (square) as well as those from Nakao and Romney (triangle), Moody et al. (circle), and Sewell and Chen (asterisk).  Note that all methods other than the proposed do not assign clusters to all actors in the network.}
\label{fratProbs}
\end{figure}

An important advantage of our proposed approach over these ad hoc or post hoc methods is the ability to compute the posterior probabilities of pairwise membership to the same cluster; that is, we can quantify the uncertainty of our hard clustering assignments.  With the MCMC output these quantities can easily be computed, and hence we can determine if the previously described results are reasonable according to our analysis.  Figure \ref{fratProbs} depicts the pairwise posterior probabilities of two actors belonging to the same cluster at week 7 (chosen for a stabilized representation of the dynamic cluster memberships).  Dark shaded regions indicate high probabilities, and light regions indicate low probabilities.  If a method estimates that two actors belong to the same cluster, then a square (our proposed method), triangle (Nakao and Romney), circle (Moody et al.), or an asterisk (Sewell and Chen) is given in the appropriate cell.  Note that all methods other than the proposed do not assign clusters to all actors in the network.  From this figure we see that there is often agreement on pairs belonging to the same cluster for most of the very high pairwise probabilities; there is also most often agreement on pairs not belonging to the same cluster for the low pairwise probabilities.  For the numerical values of the pairwise posterior probabilities for week 7 as well as for all other weeks, see the supplementary material.

\begin{figure}[!h]
\centering
\includegraphics[height=0.75\textwidth]{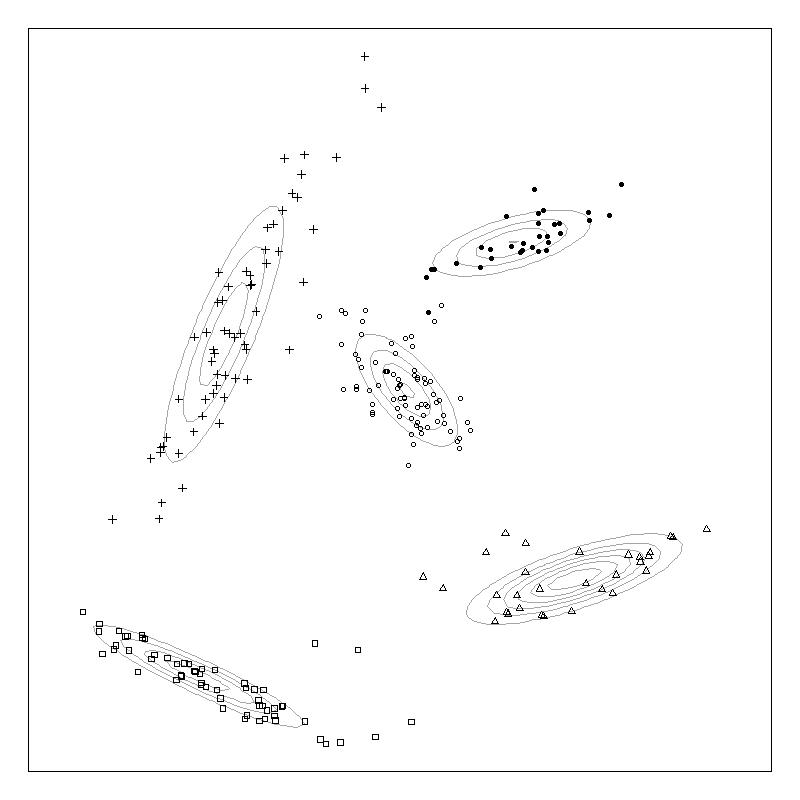}
\caption{Latent positions of all actors at all time points in the fraternity data.  The contour lines correspond to the normal distributions which characterize the five communities.  The symbols correspond to the community assignments given.}
\label{frat_overall}
\end{figure}

\begin{figure}[!h]
\centering
\begin{subfigure}[b]{0.4\textwidth}
\includegraphics[height=\textwidth]{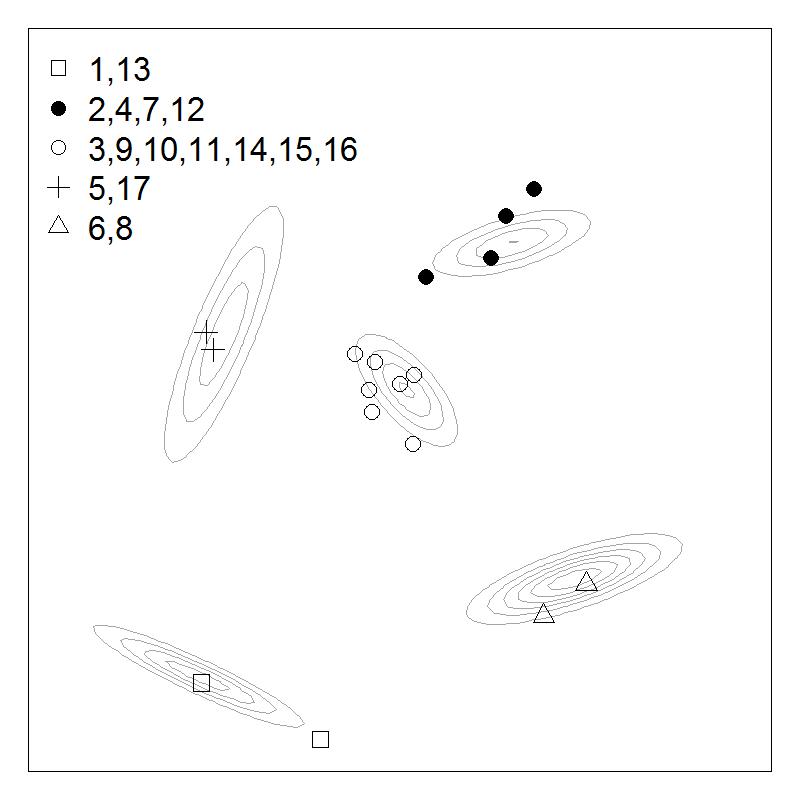}
\caption{Week 1}
\end{subfigure}
\begin{subfigure}[b]{0.4\textwidth}
\includegraphics[height=\textwidth]{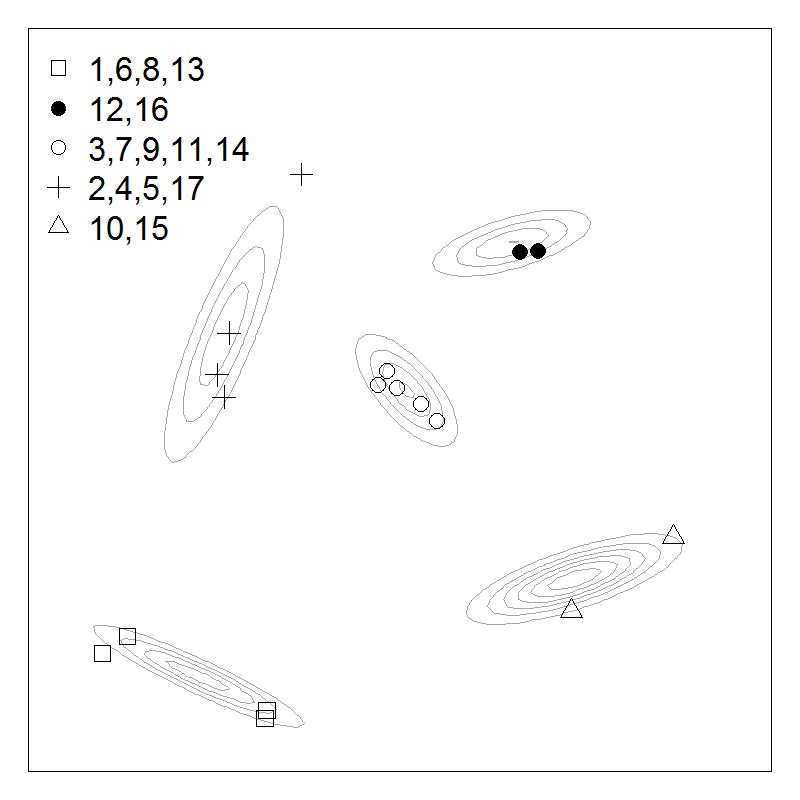}
\caption{Week 7}
\end{subfigure}\\
\begin{subfigure}[b]{0.4\textwidth}
\includegraphics[height=\textwidth]{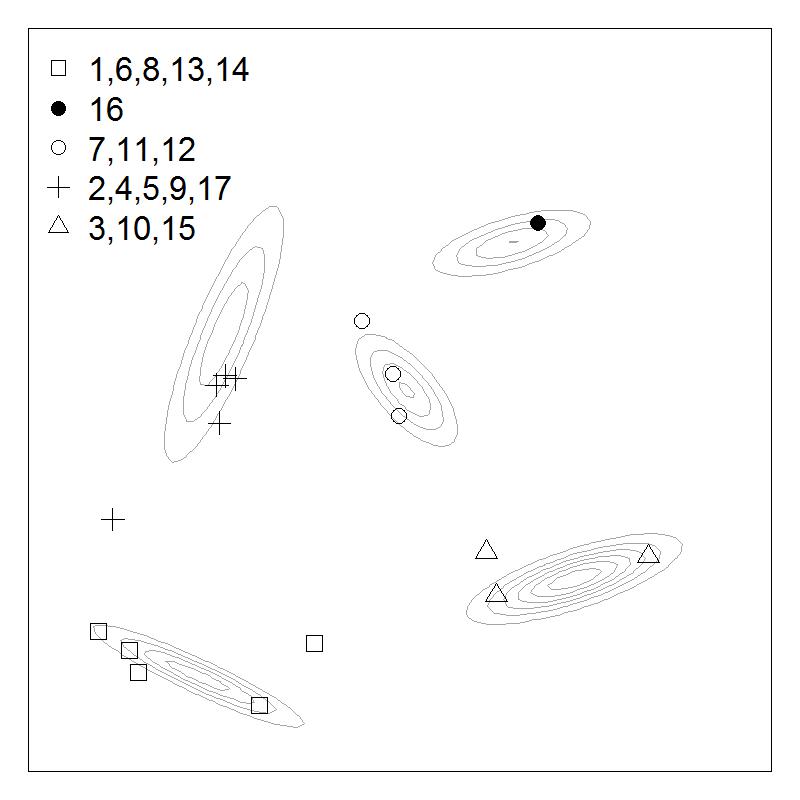}
\caption{Week 15}
\end{subfigure}
\caption{Latent positions of the fraternity data at weeks 1, 7, and 15.  The contour lines correspond to the normal distributions which characterize the five communities.  The symbols correspond to the community assignments given.}
\label{frat_snapshots}
\end{figure}

Figure \ref{frat_overall} shows the latent space with the MAP estimators of the latent positions, thus showing the overall structure of the subgroups of the network.  All actors at all time points are shown here.  Figure \ref{frat_snapshots} shows the latent positions at weeks 1, 7, and 15.  The community structure stabilized at around week 4, where it did not change at all until week 12, and only slightly until week 14.  We can characterize our five communities, referencing these groups using the shapes given in Figures \ref{frat_overall} and \ref{frat_snapshots}.  The $\Box$ community matches well with communities discovered by Nakao and Romney, Sewell and Chen, and the main community discovered by Moody et al..  Once all the members eventually joined this community within the first few weeks (none departed the community), it remained constant for the remainder of the study until student 14 joined the final week.  The \textbullet\hspace{0.1pc} community seemed to be the opposite, in that it was the most transient.  Similar to the \textbullet\hspace{0.1pc} community, the $\bigcirc$ community was also fairly transient, with many students leaving and some joining throughout the study.  The + community was characterized by students joining and remaining in the community, and in this manner was similar to the $\Box$ community.  The + community was also the most popular group in terms of rankings received, unlike the $\Box$ community which was more isolated and not very popular, and matches well with a community discovered by Sewell and Chen.  The $\triangle$ community evolved into the least popular group (until the least popular student, 16, formed his own community the last two weeks), consisting of several of those students Nakao and Romney termed ``outliers,'' and several of the students that Sewell and Chen described as having departed the main communities.

As the network was completely nascent at the first week, it is hardly a surprise that there are quite a number of actors that switch communities, especially during the beginning of the study.  Our model was able to capture this evolution of the network, unlike clustering algorithms which assume constant cluster assignments over time.  In all there were 15 transitions, 8 of which were during the first three transition periods, and 5 of which were during the last two transition periods.  This implies that the subgroup formation of the social network was fairly stable after week four, though the stability of the network faltered at the end of the semester; this last comment regarding the deterioration of the network stability also corroborates statements made by various other researchers \cite[e.g.,][]{nakao1993longitudinal,krivitsky2012rank,sewell2014analysis}.

\subsection{World trade data}
\label{WorldTradeData}
We consider world trade data with the goals of determining trade blocs and gleaning what information we can from these blocs.  We look at annual export and import data between countries during the years 1964 to 1976 (so $T=13$).  A (directed) trade relation is established from country $i$ to country $j$, i.e., $Y_{ijt}=1$, if country $i$ exports some non-negligible amount of goods to country $j$ during year $t$.  During this time, for a variety of reasons a few countries are not constant throughout, and so we only include the $n=111$ countries which exist throughout the entirety of the study period.  Thus we have thirteen $111\times111$ binary adjacency matrices.  As this is primarily a pedagogical example, we chose these years to strike a balance between a large number of time points with a large number of countries.  The data we used were obtained through the Economic Web Institute at {\it http://www.economicswebinstitute.org/worldtrade.htm}, originally obtained through the IMF Direction of Trade Yearbook.

\begin{figure}[h!]
\centering
\begin{subfigure}[b]{0.4\textwidth}
\includegraphics[height=\textwidth]{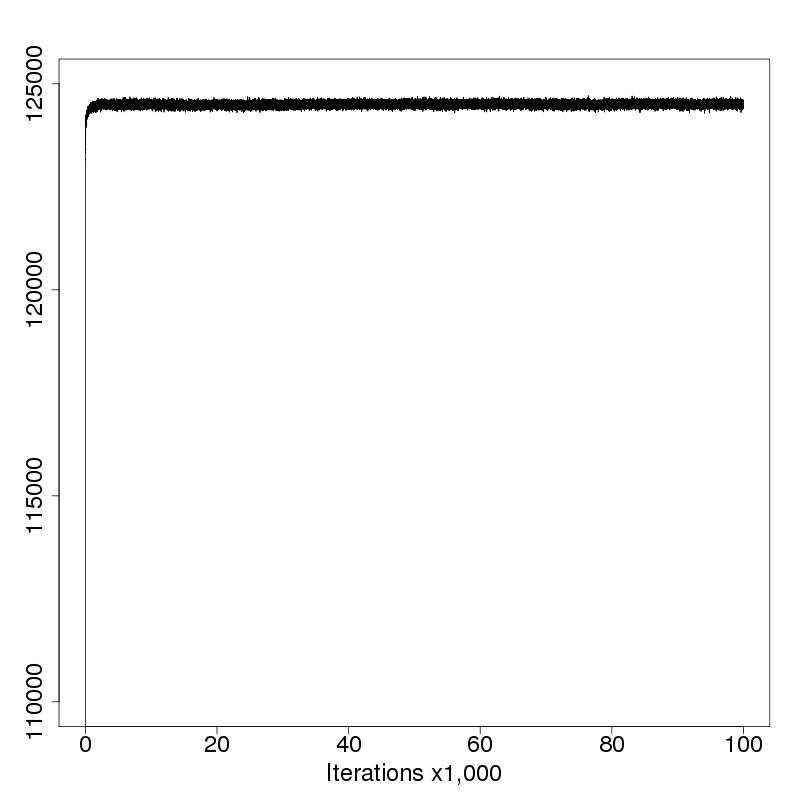}
\caption{Trace plot of the posterior value for each iteration of the MCMC algorithm.}
\label{WTTrace}
\end{subfigure}
\hspace{3pc}
\begin{subfigure}[b]{0.4\textwidth}
\includegraphics[height=\textwidth]{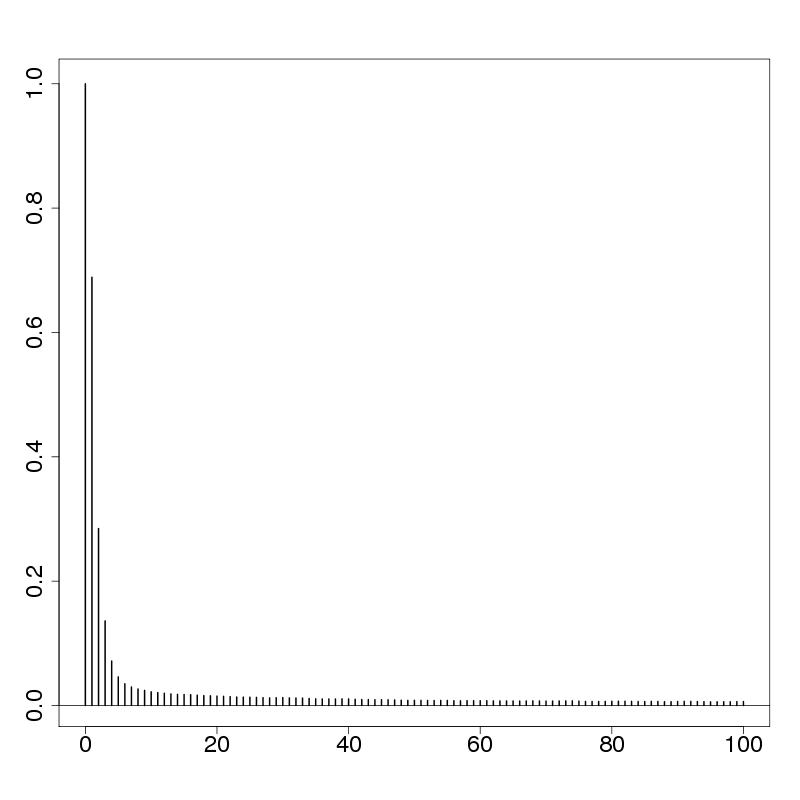}
\caption{Autocorrelation function (ACF) plot.}
\label{WTACF}
\end{subfigure}
\caption{Diagnostic plots for MCMC estimation corresponding to the world trade data.}
\end{figure}

To detect trade blocs within the binary trade relations data, we implemented both the distance model and the projection model, letting $G$ take values from 2 to 9.  Using the procedure described in Sections \ref{NumberOfClusters} and \ref{BICsimstudy}, we selected the projection geometry with four clusters;  the BIC values for the projection model ranged from $-35,838$ to $-34,448$.  Figure \ref{WTTrace} provides a trace plot of the posterior value of all 100,000 samples, and Figure \ref{WTACF} provides the ACF plot.  From these we see evidence of convergence and good mixing.  Geweke's diagnostic test yielded a $p$-value of 0.210 using a burn in of 35,000, implying convergence.

\begin{figure}[h!]
\centering
\includegraphics[height=0.5\textwidth]{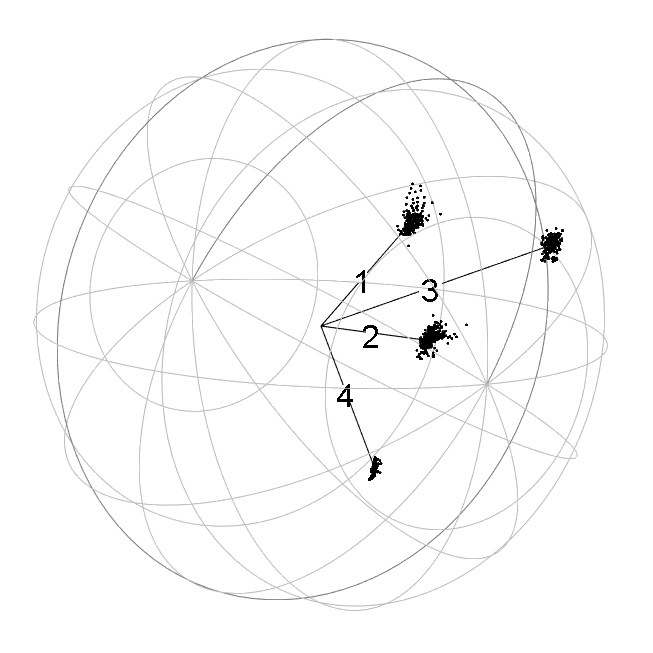}
\caption{Estimates of latent locations (plotting the unit vectors indicating direction and ignoring the magnitude of the vectors that correspond to individual effects) of countries in the international export/import data.  The four communities have been labeled along segments from the origin to the communities' centers.}
\label{worldTradeFig}
\end{figure}

Figure \ref{worldTradeFig} shows the posterior mode of the latent positions of all countries at all time points ($nT$ points plotted), where the four communities have been labeled along segments from the origin to the communities' centers.  For ease of viewing we have plotted the countries based only on their directional unit vectors, disregarding the magnitudes of the vectors which correspond to the individual effects.

Most of the blocs are relatively densely interconnected, as seen in Table \ref{WTDensities}.  The exception is Bloc 1,
a global trade bloc with nations representing all inhabited continents, which is loosely interconnected.  This community is also the most transitory, as seen in Table \ref{WTBetas} which gives the estimated values of $\bbeta_0$ and $\bbeta_h$, $h=1,\ldots,4$.  It is intuitive that these two things should coincide, in that trade blocs that are not actively trading with each other should be more likely to lose member nations to other trade blocs.  Bloc 2 is the largest bloc averaging 49 nations per year, and involves with very few exceptions only eastern hemisphere nations, indicating that geography may be playing a role in the formation of trade blocs.  Bloc 3 consists of the U.S.S.R., several eastern European countries, and most of Latin America.  This gives quantitative evidence in favor of claims of close ties between U.S.S.R. and Latin America and the Soviet influence in the western hemisphere \citep[e.g.,][]{blasier1988giant}.  Bloc 4 is a community that is indicative of a very interesting vestigial effect from French colonization.  Of the countries that belonged to bloc 4, France and her former colonies constitute $2/3$ of them.  French colonial policy required her colonies to import only from or through France, export only to France, and to ship using French vessels \citep{grier1999colonial}.  That France and her former colonies behave similarly as participants in world trade gives evidence that colonial policy established a longer term trend.

\begin{table}[t]
\centering
\begin{tabular}{r|rrrr}
 & 1 & 2 & 3 & 4 \\
  \hline
1 & 0.10 & 0.14 & 0.08 & 0.07 \\
  2 & 0.14 & 0.25 & 0.15 & 0.12 \\
  3 & 0.08 & 0.15 & 0.24 & 0.05 \\
  4 & 0.07 & 0.12 & 0.05 & 0.19 \\
\end{tabular}
\caption{Densities within each of the four communities and between each community, averaged over all time points.  These densities are computed by dividing the total number of edges by the total possible number of edges.}
\label{WTDensities}
\end{table}

\begin{table}[t]
\centering
\begin{tabular}{r|rrrr}
&\multicolumn{4}{c}{$g$}\\
 & 1 & 2 & 3 & 4 \\
  \hline
0 & 0.305 & 0.394 & 0.217 & 0.084 \\
  1 & 0.952 & 0.034 & 0.013 & 0.001 \\
  2 & 0.002 & 0.983 & 0.009 & 0.006 \\
  3 & 0.011 & 0.005 & 0.983 & 0.002 \\
  4 & 0.004 & 0.004 & 0.004 & 0.989 \\
\end{tabular}
\caption{Estimates of initial clustering parameter (first row) and transition parameters (last four rows).}
\label{WTBetas}
\end{table}

\section{Discussion}
\label{Discussion}
Community detection is an important topic in network analysis.  We have extended the commonly used distance and projection latent space models to incorporate clustering of dynamic network data, utilizing the temporal information to build the model.  This model can handle directed or undirected dynamic network data, and can also be used to model a wide range of weighted network data.  We have also given the first, to our knowledge, clustering model corresponding to the projection model in \cite{hoff2002latent}, \cite{durante2014nonparametric}, and others.  This model also can handle directed or undirected dynamic network data, and the VB algorithm we have described provides computationally fast estimation of the model.

While the VB algorithm using the projection model for binary networks is relatively fast, the corresponding Gibbs sampler we have also implemented is time intensive for larger networks, as seen in Figure \ref{compTime}.  However, this burden could potentially be alleviated by adapting the likelihood approximation method first derived by \cite{raftery2012fast} for binary networks.  For the distance model, we expect that creating a VB algorithm would be non-trivial and context specific; we therefore leave that for future research.

In this paper we have discussed a method of selecting the number of clusters and the latent space geometry.  However, a difficult topic we have not yet addressed is the selection of the dimension of the latent space.  \cite{durante2014nonparametric} developed a non-parametric approach to this problem in a simpler setting, which may inspire similar type strategies to select the dimensionality of the latent space in our context.  A very useful area of future research then would be to construct a unifying model selection method to determine the latent space geometry, the dimension of the latent space, and the number of clusters.

One last comment is that the clustering models that have been proposed are based on the assumption that actors within a cluster are more likely to form edges than actors in different clusters.  While this is, we expect, the most common context, there may be certain scenarios in which this is not the case.  Instead there may be varying roles that the actors can take on, and these roles do not necessitate that each role is well interconnected, that is, actors in the same community may not be densely connected to each other.  In such a case a blockmodel approach would be more appropriate to modeling the data.

\bibliographystyle{ba}
\bibliography{DNC}

\section*{Acknowledgements}
The authors thank the editor, the associate editor, and the referees for valuable suggestions.
This work was supported in part by National Science Foundation grants DMS-1106796 and DMS-1406455.

\end{document}